\documentclass[final]{aipproc}
\layoutstyle{6x9}
\usepackage{amssymb}

\newcommand\Alfven{Alfv\'en }
\begin{document}

\title{Dissipation-Scale Turbulence in the Solar Wind}


\author{G.~G. Howes}{
address={Department of Astronomy, University of California, Berkeley,
CA, USA.}  }
\author{S.~C. Cowley}{
address={Department of Physics and Astronomy, UCLA, Los Angeles, CA,
USA.}  }
\author{W. Dorland}{
address={Department of Physics and Center for Scientific Computing and
Mathematical Modeling, University of Maryland, College Park, MD, USA.}
}
\author{G.~W. Hammett}{
address={Princeton Plasma Physics Laboratory, Princeton, NJ, USA.}  }
\author{E. Quataert}{
address={Department of Astronomy, University of California, Berkeley,
CA, USA.}  }
\author{A.~A. Schekochihin}{
address={King's College, Cambridge, UK and Department of Physics,
Imperial College, London, UK.}  }

\begin{abstract}
We present a cascade model for turbulence in weakly collisional
plasmas that follows the nonlinear cascade of energy from the large
scales of driving in the MHD regime to the small scales of the kinetic
\Alfven wave regime where the turbulence is dissipated by kinetic
processes.  Steady-state solutions of the model for the slow solar
wind yield three conclusions: (1) beyond the observed break in the
magnetic energy spectrum, one expects an exponential cut-off; (2) the
widely held interpretation that this dissipation range obeys power-law
behavior is an artifact of instrumental sensitivity limitations; and,
(3) over the range of parameters relevant to the solar wind, the
observed variation of dissipation range spectral indices from $-2$ to $-4$
is naturally explained by the varying effectiveness of Landau damping,
from an undamped prediction of $-7/3$ to a strongly damped index around
$-4$.

\end{abstract}

\maketitle

%%%%%%%%%%%%%%%%%%%%%%%%%%%%%%%%%%%%%%%%%%%%
%% MAINMATTER
%%%%%%%%%%%%%%%%%%%%%%%%%%%%%%%%%%%%%%%%%%%%
%=============================================================================
\section{Introduction}

One of the principal measurements in the study of solar wind turbulence
is the magnetic field fluctuation frequency spectrum derived from
\emph{in situ} satellite measurements. At 1~AU, the one-dimensional
energy spectrum in spacecraft-frame frequency typically shows, for low
frequencies, a power law spectrum with slope of -5/3, suggestive of a
Kolmogorov-like inertial range \citep{Coleman:1968,Goldstein:1995}; a
spectral break is typically observed at around 0.4 Hz, with a steeper
power law at higher frequencies, often denoted the dissipation range
in the literature, with a spectral index that varies from -2 to -4
\citep{Leamon:1998a,Smith:2006}. The general consensus is that the
-5/3 portion of the spectrum is the inertial range of an MHD turbulent
cascade, but the dynamics responsible for the break and steeper
portion of the spectrum is not well understood. Various explanations
for the location of the break in the spectrum have been proposed: that
it is coincident with the proton cyclotron frequency in the plasma
\citep{Goldstein:1994,Leamon:1998a,Gary:1999}, or that
the fluctuation length scale has reached either the proton Larmor
radius \citep{Leamon:1998b,Leamon:1999} or the proton inertial length
\citep{Leamon:2000,Smith:2001b}.  The steepening of the spectrum at
higher wavenumbers has been attributed to proton cyclotron damping
\citep{Coleman:1968,Goldstein:1994,Leamon:1998a,Gary:1999},
Landau damping of kinetic \Alfven waves
\citep{Leamon:1998b,Leamon:1999,Leamon:2000}, or the
dispersive nature of whistler waves \citep{Stawicki:2001}.

To unravel the underlying physical mechanisms at work in the solar
wind requires an understanding of turbulence in weakly collisional,
magnetized plasmas. Early theories of MHD turbulence proposed an
isotropic cascade of turbulent energy
\citep{Iroshnikov:1963,Kraichnan:1965}, but numerical simulations
\citep{Shebalin:1983} demonstrated
an inherent anisotropy in the presence of a mean magnetic field. An
evolving anisotropic theory  
\citep{Montgomery:1982,Shebalin:1983,Higdon:1984a,Sridhar:1994,Goldreich:1995}
has emerged which rests upon two central hypotheses: the Kolmogorov
hypothesis of locality in wavenumber space \citep{Kolmogorov:1941},
and the conjecture that in strong turbulence the linear wave periods
maintain a critical balance with the nonlinear turnover timescales.
The anisotropic nature of the turbulence means the frequency for
nonlinear energy transfer is dominated by the perpendicular
wavenumber, $k_\perp v_\perp$, where $\perp$ denotes the component
perpendicular to the mean magnetic field. Assuming balanced turbulence
with equal Els\"asser energy fluxes in either direction along the mean
field, the one-dimensional magnetic energy spectrum in the MHD regime
scales as $E_B(k_\perp) \propto k_\perp^{-5/3}$ and critical balance
implies a scale-dependent anisotropy with $k_\parallel \propto
k_\perp^{2/3}$ \citep{Goldreich:1995}. In the regime of electron MHD
(EMHD) \citep{Kingsep:1990}, one obtains $E_{B}(k_\perp) \propto
k_\perp^{-7/3}$ and $k_\parallel \propto k_\perp^{1/3}$
\citep{Biskamp:1999,Cho:2004}.

Are observations of turbulence in the solar wind consistent with these
theoretical predictions?  The energy in turbulent fluctuations is
observed to be anisotropic \citep{Matthaeus:1990} with $k_\perp >
k_\parallel$ in the slow solar wind at scales of  $k_\perp
\rho_i \sim 10^{-3}$ \citep{Dasso:2005}, where $\rho_i$ is the proton 
Larmor radius; this appears consistent with the prediction of a scale
dependent anisotropy leading to nearly perpendicular wavevectors
$k_\perp \gg k_\parallel$ at small scales. The imbalance between
anti-sunward and sunward Els\"asser spectra can reach nearly two
orders of magnitude in the fast wind, while the slow wind has a much
smaller imbalance, from a factor of a few to approximate equality
\citep{Tu:1990a,Grappin:1990}.  Thus, we believe the aforementioned
theory of MHD turbulence to be relevant to the dynamics in the slow
solar wind.

Although the large scales at which the turbulence is driven may be
adequately described by MHD, the turbulent fluctuations at the
small-scale end of the inertial range often have parallel wavelengths
smaller than the ion mean free path; therefore, a kinetic description
of this weakly collisional plasma is required to capture the turbulent
dynamics. The slow, fast, and entropy modes are damped in a warm,
collisionless plasma \citep{Barnes:1966}; the \Alfven wave cascade,
however, is undamped until it reaches the ion Larmor radius, $k_\perp
\rho_i \sim 1$ \citep{Quataert:1998,Schekochihin:2007}. For a sufficiently 
large inertial range, wavevectors at this scale become nearly
perpendicular with $k_\perp \gg k_\parallel$; thus, frequencies remain
low compared to the ion cyclotron frequency $\omega < \Omega_i$, the
nonlinear cascade to yet smaller scales is composed of kinetic \Alfven
waves, and Landau damping by the ions and electrons can effectively
dissipate the turbulence.  The dynamics in this regime optimally
described by a low-frequency limit of kinetic theory called
gyrokinetics \citep{Howes:2006,Schekochihin:2007}. Here we present a
model aimed at following the nonlinear cascade of magnetic energy from
fluid to kinetic scales while accounting for the kinetic dissipation of
the turbulence.

%=============================================================================
\section{Analytical Model}
Consider a homogeneous magnetized plasma with a mean magnetic field of
magnitude $B_0$ that is stirred isotropically at an outer scale
wavenumber $k_0$ with velocity $v_0$.  We write the magnetic field
fluctuations in velocity units, $b_k \equiv \delta B_\perp (k_\perp)
/\sqrt{4 \pi n_i m_i}$. The frequency of nonlinear energy transfer for
Alfv\'enic fluctuations at a given perpendicular wavenumber is
estimated to be $\omega_{nl} \sim k_\perp v_k = k_\perp b_k$. Assuming
the locality of nonlinear interactions in wavenumber space and a
constant energy cascade rate $\epsilon$, the one-dimensional magnetic
energy spectrum in the regime $k_\perp \rho_i \ll 1$ is given by
\begin{equation}
E_{B}(k_\perp) = \frac{b_k^2}{k_\perp} = C_{1m} \epsilon^{2/3} k_\perp^{-5/3},
\label{eq:ekmhd}
\end{equation}
where $C_{1m}$ is a dimensionless constant of order unity.  The
frequency of nonlinear energy transfer is
\begin{equation}
\omega_{nl} = C_{2m} \epsilon^{1/3} k_\perp^{2/3},
\label{eq:wnlmhd}
\end{equation}
where $C_{2m}$ is another order unity constant; the parallel
wavenumber can be determined by applying the critical balance
conjecture, setting the linear \Alfven wave frequency equal to the
nonlinear frequency $\omega=\omega_{nl}$.

In the kinetic \Alfven wave regime $k_\perp \rho_i \gg 1$, the
dynamics are governed by the equations of Electron Reduced MHD
\citep{Schekochihin:2007}, with characteristic fluctuations
$v_k = \pm  b_k  k_\perp \rho_i/\sqrt{\beta_i + 2/(1+T_e/T_i)}$.
Applying the same procedure for this regime yields the
one-dimensional magnetic energy spectrum
\begin{equation}
E_{B}(k_\perp) = \frac{b_k^2}{k_\perp} = C_{1k} \epsilon^{2/3} 
\frac{[\beta_i + 2/(1+T_e/T_i)]^{1/3}} {\rho_i^{2/3}} k_\perp^{-7/3},
\label{eq:ekkaw}
\end{equation}
and the nonlinear frequency
\begin{equation}
\omega_{nl} = C_{2k} \epsilon^{1/3} \frac{\rho_i^{2/3}}
{[\beta_i + 2/(1+T_e/T_i)]^{1/3}}  k_\perp^{4/3}.
\label{eq:wnlkaw}
\end{equation}

A continuity equation for the magnetic energy per unit mass at each
wavenumber $b_k^2$ can be written as \citep{Batchelor:1953}
\begin{equation}
\frac{\partial b_k^2}{\partial t} = 
-\frac{\partial \epsilon(k_\perp) }{\partial \ln k_\perp} 
+ S(k_\perp) - 2 \frac{\gamma(k_\perp)}{\omega(k_\perp)} 
\omega_{nl}(k_\perp) b_k^2,
\label{eq:modelcont}
\end{equation}
where the three terms on the right-hand side are the energy flux through
wavenumber space, a source term, and a damping term. The energy
cascade rate is modeled by
\begin{equation}
\epsilon (k_\perp)= 
k_\perp b_k^3 \left[
C_{1m}^{-3} + \frac{C_{1k}^{-3} (k_\perp \rho_i)^2}
{\beta_i + 2/(1+T_e/T_i)} \right]^{1/2}
\end{equation}
and the nonlinear frequency by
\begin{equation}
\omega_{nl} (k_\perp)= 
k_\perp b_k  \left[ \frac{C_{2m}^2}{C_{1m}}
 +\frac{C_{2k}^2}{C_{1k}}  \frac{(k_\perp \rho_i)^2}
{\beta_i + 2/(1+T_e/T_i)} \right]^{1/2}.
\end{equation}
In the damping term, $\gamma/\omega$ is determined from the linear
gyrokinetic dispersion relation \citep{Howes:2006}.  The order unity
constants are taken to be $C_{1m}=C_{1k}=2.5$ and $C_{2m}=C_{2k}=2.2$
based on numerical simulation, as in \citet{Quataert:1999}.

%=============================================================================
\section{Results}
\label{sec:results}
\begin{figure}
\hbox{\resizebox{\textwidth}{!}{
\resizebox{\textwidth}{!}{\includegraphics{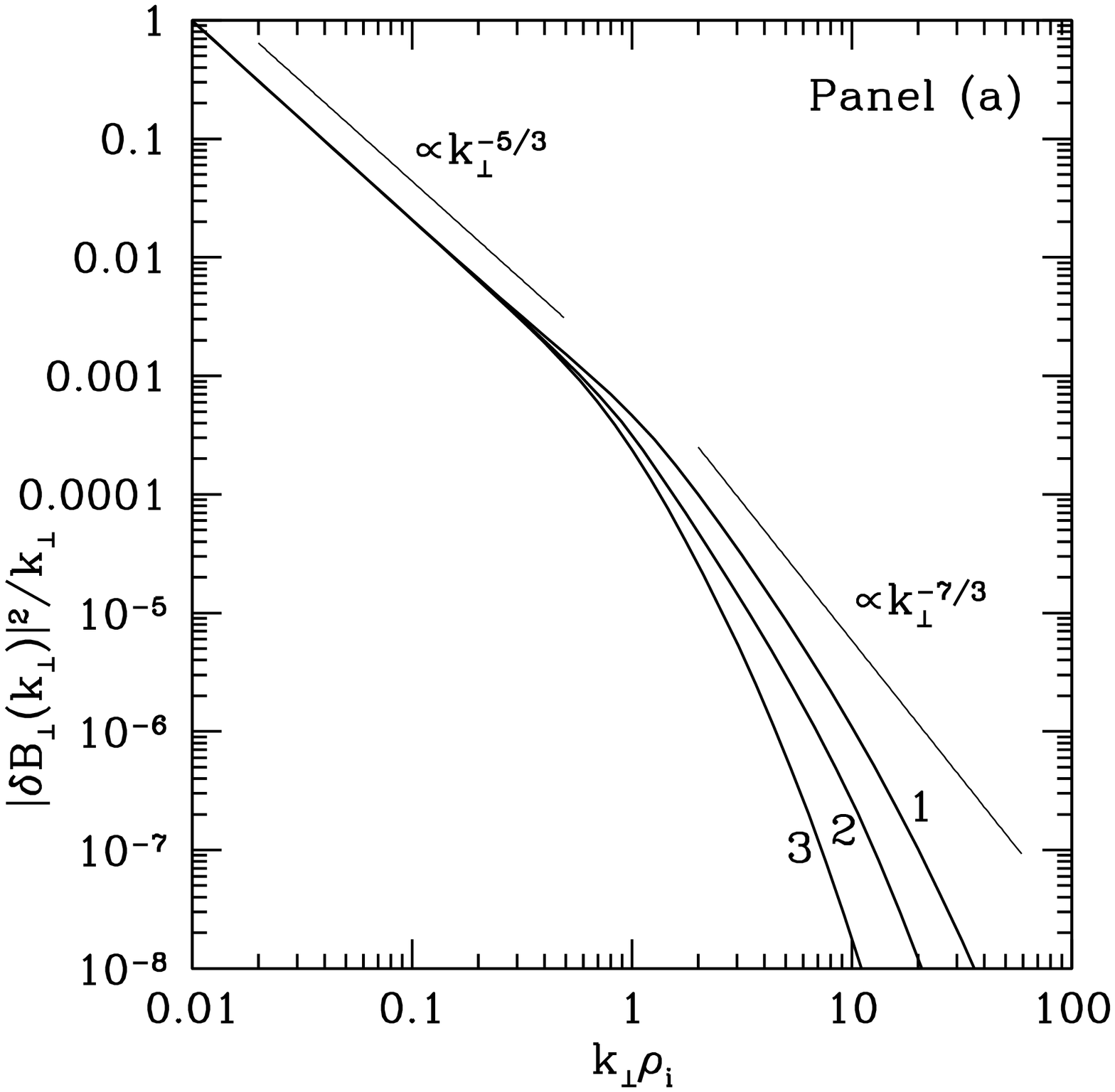}} \hfill
\resizebox{\textwidth}{!}{\includegraphics{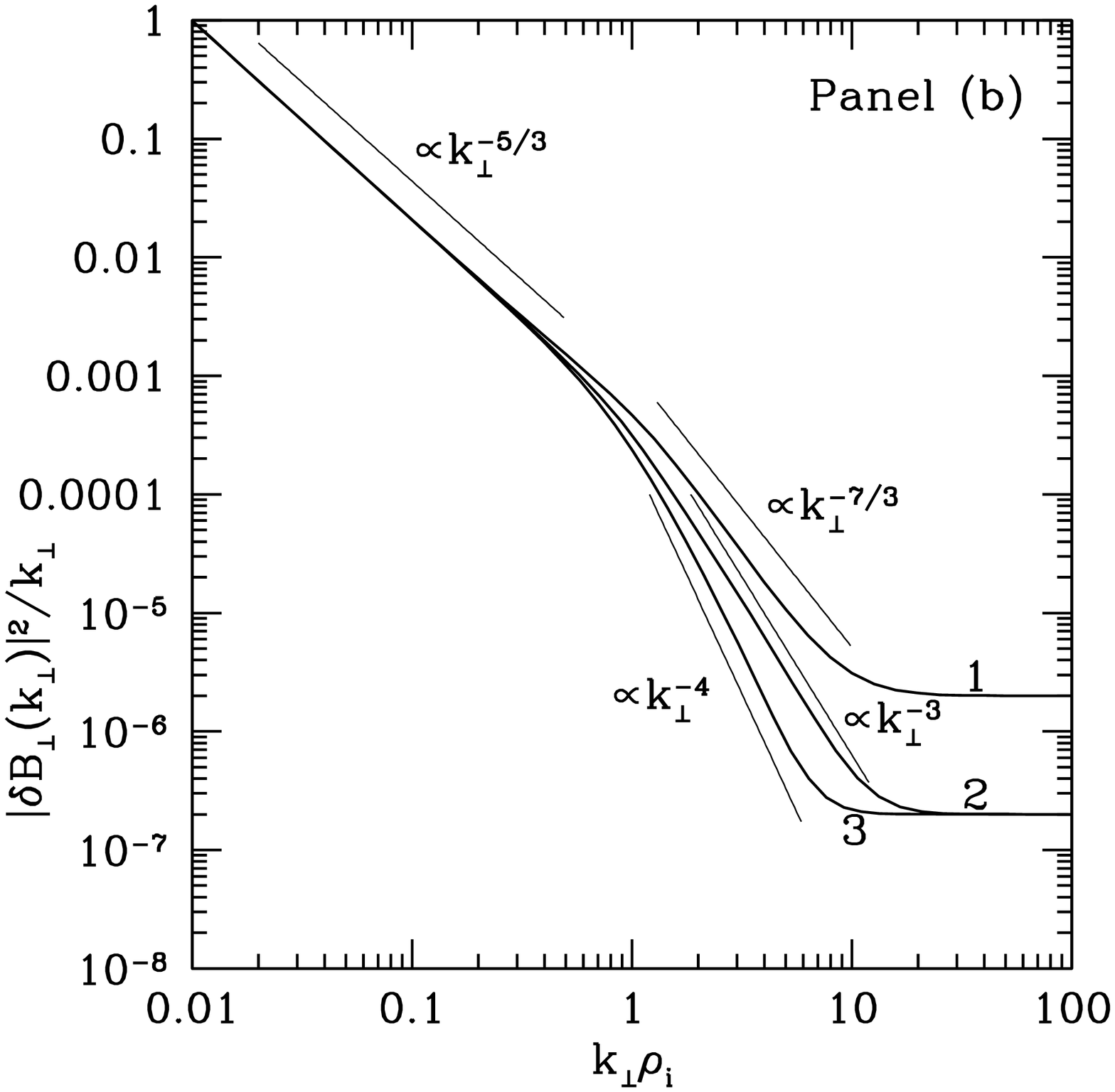}}}}
  \caption{\label{fig:diss_range}One-dimensional magnetic energy
  spectra for three gyrokinetic models: (1) $\beta_i=0.5$, $T_i/T_e=3$,
  (2) $\beta_i=3$, $T_i/T_e=0.6$, (3) $\beta_i=0.03$, $T_i/T_e=0.175$.
  All models use $k_0 \rho_i=3\times 10^{-5}$.  Panel (a) shows
  that all three spectra demonstrate a dissipative roll-off with a
  variation of spectral indices in the range $k_\perp \rho_i>1$. Panel
  (b) adds a constant magnetometer sensitivity limit to each spectrum,
  yielding dissipation range spectra that more closely resemble power
  laws with a range of slopes from $-7/3$ to $-4$ .}
\end{figure}

The model given by equation \eqref{eq:modelcont} is solved numerically
to obtain a steady state magnetic energy spectrum for a given set of
the parameters ion plasma beta $\beta_i$, ion to electron temperature
ratio $T_i/T_e$, and isotropic driving scale $k_0 \rho_i$. Panel (a)
of Figure~\ref{fig:diss_range} presents the solutions for three cases
chosen to sample the observed parameter range in the solar wind
\citep{Leamon:1998a,Newbury:1998,Howes:2007}: (1) $\beta_i=0.5$,
$T_i/T_e=3$; (2) $\beta_i=3$, $T_i/T_e=0.6$; and (3) $\beta_i=0.03$,
$T_i/T_e=0.175$.  All models use $k_0\rho_i=3\times 10^{-5}$.  In the
absence of dissipation, analytical theory predicts spectral indices of
$-5/3$ in the MHD regime and $-7/3$ in the kinetic \Alfven wave
regime; with the damping rate artificially set to zero, this model
indeed recovers these results (not shown).  Damping at $k_\perp \rho_i
\gtrsim 1$ is sufficient to cause each spectrum in panel (a) to 
fall off more steeply than the undamped prediction of $-7/3$. The
steady-state spectra obtained here clearly demonstrate the exponential
roll-off characteristic of dissipation \citep{Li:2001}.

Observations of the magnetic fluctuation spectra at higher
wavenumbers than the spectral break are widely interpreted to behave
like a power law rather than an exponential decay.  We suggest here
that the power-law appearance of the spectrum in this range is an
effect of limited magnetometer sensitivity; this sensitivity limit can
be clearly seen in Figure~6 of \citet{Leamon:1998a} at the high
wavenumber end of the spectrum.  The noise floor of a fluxgate
magnetometer at frequencies $f> 1$ Hz is constant in units of
nT/$\sqrt{\mbox{Hz}}$ \citep{Bale:2007pc}, so we mock up the
instrumental noise by specifying a constant background value of the
one-dimensional energy spectrum. We choose the noise floor to be
approximately two to three orders of magnitude below the spectrum
value at the break. Panel (b) of Figure~\ref{fig:diss_range} adds a
constant sensitivity level at two (spectrum 1) or three (spectra 2 and
3) orders of magnitude below the spectrum value at $k_\perp \rho_i
=1$.  The behavior of each spectrum in panel (b) in the range $k_\perp
\rho_i >1$ more closely resembles a power law than the exponential
roll-off in the noiseless spectra; the steady-state solutions are
well-fit by power laws with spectral indices $-7/3$, $-3$, and $-4$.  In
summary, the instrumental sensitivity limit is crucial in interpreting
measured magnetic fluctuation spectra, and may produce spectra that
appear to obey a power-law scaling even though the underlying spectrum
is actually rolling off exponentially.

The spectral index in the dissipation range is observed to vary from
$-2$ to $-4$ \citep{Leamon:1998a,Smith:2006}. Figure~\ref{fig:diss_range}
shows that, over the range of the plasma parameters $\beta_i$ and
$T_i/T_e$ measured in the solar wind, this variation can naturally be
explained by the varying effectiveness of the damping of kinetic
\Alfven waves via the Landau resonance.  If Landau damping is negligible,
the spectral index is expected to give a value of $-7/3$, close to the
observed upper limit; over the range of parameters relevant to the
solar wind, this cascade model gives a lower limit to the spectral
index of about $-4$, for example spectrum~3, consistent with
observations.  Hence, the varying effectiveness of Landau damping is
sufficient to explain the observed variation of spectral indices, with
the break occurring at the ion Larmor radius.

%=============================================================================
\section{Conclusion}
The physical mechanisms responsible for the spectral break and steeper
dissipation range of the magnetic energy spectrum observed in the
solar wind have not been conclusively identified. This paper presents
a turbulent cascade model constructed to follow the magnetic
fluctuation energy from the large scales in the MHD regime down to the
small scales in the kinetic \Alfven wave regime, accounting for
dissipation by kinetic processes. Due to the inherent anisotropy of
MHD turbulence, the turbulence remains low frequency $\omega \ll
\Omega_i$ and is optimally described by gyrokinetics. This picture of
balanced, low-frequency turbulence is relevant to the slow solar wind.

The nonlinear cascade model given by \eqref{eq:modelcont} using the
gyrokinetic damping rates is solved numerically to find steady-state
solutions as presented in Figure~\ref{fig:diss_range}. The cascade
model predicts that, for wavenumbers above the break in the magnetic
fluctuation energy spectrum, the spectrum undergoes a slow exponential
cut-off. We argue that the widespread interpretation that this
dissipation range shows power-law behavior is an artifact of limited
magnetometer sensitivity. Over the range of parameters $\beta_i$ and
$T_i/T_e$ measured in the solar wind, the varying strength of Landau
damping naturally reproduces the observed variation of dissipation
range spectral indices from $-7/3$ to $-4$, with the spectral break
occurring at the scale of the ion Larmor radius.

This model assumes that linear damping rates are relevant for
turbulent fluctuations that are nonlinearly cascaded to smaller scales
on the timescale of one wave period.  Nonlinear gyrokinetic
simulations of the turbulent cascade in the transition to the kinetic
\Alfven wave regime are necessary to judge the validity of this
assumption.  This cascade model can be used as a tool to connect
nonlinear numerical simulations to observations of turbulence in the
solar wind. Further work to examine the importance of the proton
cyclotron resonance in dissipation of solar wind turbulence is
underway \citep{Howes:2007}.

\begin{theacknowledgments}
Thanks to S. Bale for helping to apply this work to the solar wind.
This work was supported by the DOE Center for Multiscale Plasma
Dynamics, Fusion Science Center Cooperative Agreement ER54785 and the
hospitality of the Aspen Center for Physics.
\end{theacknowledgments}

\bibliographystyle{aipproc} 
\bibliography{abbrev2,solwind,turbulence,gyro,gen_plasma}

\end{document}